\documentstyle[psfig,aps,prl,twocolumn]{revtex}

\begin{document}
\draft

\wideabs{

\author{Michael K\"ohl, Theodor W.~H\"ansch, and Tilman
Esslinger}
\address{Sektion Physik,
Ludwig-Maximilians-Universit\"at, Schellingstr.\ 4/III, D-80799 Munich, Germany and\\
Max-Planck-Institut f\"ur Quantenoptik, D-85748 Garching, Germany}
\title{Continuous detection of an atom laser beam}
\date{Submitted September 4, 2001}
\maketitle

\begin{abstract}
\noindent We have demonstrated a detection scheme for atom laser
beams that allows for a continuous measurement of the atom density
and readout of the data in real-time. The atoms in the atom laser
beam are transferred locally  from the lower to the upper
hyperfine ground state of $^{87}$Rb by coherent coupling and are
subsequently detected by an absorption measurement. The detection
is non-destructive to the Bose-Einstein condensate in the magnetic
trap and the atom laser beam remains unaffected outside the
detection region.

\end{abstract}

}

\noindent

The possibility to extract coherent atomic beams from
Bose-Einstein condensates \cite{atomlasers} has stimulated
enormous theoretical and experimental interest in this field.
Very recently, the temporal coherence of atom laser beams was
measured \cite{Koehl2001a} and their transverse divergence
investigated \cite{Lecoq2001a}. For studies of atom laser beams
only absorption imaging has been employed so far. This technique
is destructive since the atoms scatter near-resonant photons and
it therefore allows only for single shot measurements. The
technique of phase-contrast imaging \cite{Andrews1996a}, where the
atoms scatter light far off resonance, is almost non-destructive
and several repetitive measurements of a single Bose-Einstein
condensate have been obtained. However, phase contrast imaging
requires large atomic densities. It is therefore only suitable
for imaging Bose-Einstein condensates in the magnetic trap but
not for atom laser beams which usually have two to three orders
of magnitude lower densities. Also, the repetition rate for phase
contrast imaging is typically a few hundred Hz which limits the
temporal resolution to values larger than a millisecond.

Besides being non-continuous there is another major drawback of
the existing imaging techniques: the time delay between the
actual measurement and the readout of the data. For slow-scan CCD
cameras this is typically a few hundred milliseconds and hinders
the realization of experiments where a measured quantity, e.g. the
optical density at a certain position, is fed back onto the
condensate.

\begin{figure}
\centerline{\psfig{file=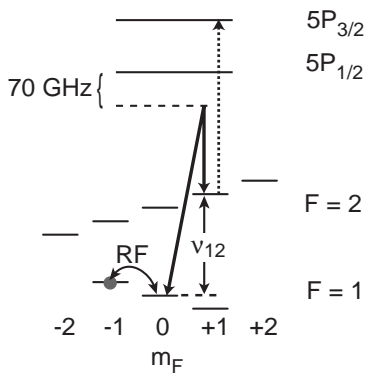,angle=0,width=0.7\columnwidth}}
\caption{Level scheme of $^{87}$Rb and the transitions involved
in the detection scheme. The Bose-Einstein condensate is formed
in the $|F=1,m_F=-1\rangle$ state and the atom laser is output
coupled by an  RF transition into the $|F=1,m_F=0\rangle$ state,
with vanishing magnetic moment. The $|F=1,m_F=0\rangle$ and the
$|F=2,m_F=1\rangle$ state are coherently coupled by a two-photon
Raman transition, which is detuned by 70\,GHz from the D$_1$ line
(solid arrow). Atoms in the F=2 ground state are detected by an
absorption measurement on the D$_2$ line using the
$F=2\rightarrow F^\prime =3$ cycling transition (dotted arrow).}
\label{fig1}
\end{figure}

We have demonstrated a continuous detection of an atom laser beam
with readout of the data in real-time. The detection scheme
employs a position-dependent coherent coupling between the
hyperfine ground states $F=1$ and $F=2$ of $^{87}$Rb. Atoms in the
atom laser beam are transferred locally from the
$|F=1,m_F=0\rangle$ into the $|F=2,m_F=1\rangle$ hyperfine ground
state, in which they are detected by an absorption measurement
(see Fig. \ref{fig1}).

The experimental procedure for the production of continuous atom
laser beams has been described previously \cite{Bloch1999a}.
Briefly, we load about $10^9$ atoms in the $|F=1,m_F=-1\rangle$
hyperfine ground state into a magnetic Quadrupole-Ioffe
configuration trap (QUIC-trap) \cite{Esslinger1998a}. The
trapping frequencies are $\omega_\perp=2\pi\times 110$\,Hz in the
radial and $\omega_z=2 \pi \times 14$\,Hz in the axial direction.
In the magnetic trap we perform forced evaporative cooling for
about 30\,s after which  Bose-Einstein condensates of $5\times
10^5$ atoms are obtained. Continuous atom laser beams are
extracted from the condensate by employing a weak, monochromatic
radio frequency field that induces locally spinflip transitions of
condensed atoms into the magnetically untrapped state
$|F=1,m_F=0\rangle$. Atoms in this state are accelerated by
gravity, propagate downward and form a collimated beam.

After a dropping distance of about 500\,$\mu$m the atoms enter a
region of focused, co-propagating laser beams which have a
cylindrical waist with 1/e$^2$-radii of 26\,$\mu$m and
500\,$\mu$m in the vertical and horizontal direction,
respectively. Two of these laser light fields couple the
$|F=1,m_F=0\rangle$ state and the $|F=2,m_F=1\rangle$ state by a
two photon hyperfine Raman transition (see Fig. \ref{fig1})
\cite{Bloch2001a}. The resonance condition for this transition is
given by the frequency difference between the Raman laser light
fields $\nu_{12}$ and the local magnetic field, which induces a
spatially dependent Zeeman shift for the $|F=2,m_F=1\rangle$
state. The Raman lasers are detuned by 70\,GHz to the red of the
5S$_{1/2}$\,$\rightarrow$\,5P$_{1/2}$ transition (D$_1$-line,
795\,nm) and the power needed to achieve 100\% transfer between
the different hyperfine levels is for our geometry 1\,mW.
Additionally, laser light resonant with the $F=2 \rightarrow
F^{\prime}=3$ transition of the D$_2$-line
(5S$_{1/2}$\,$\rightarrow$\,5P$_{3/2}$, 780\,nm) is superimposed
to the pair of Raman laser beams. As soon as the atoms are
transferred into the $|F=2,m_F=1\rangle$ state they absorb light
out of this beam and its extinction is a measure of the density
of the atom laser beam.

\begin{figure}
\centerline{\psfig{file=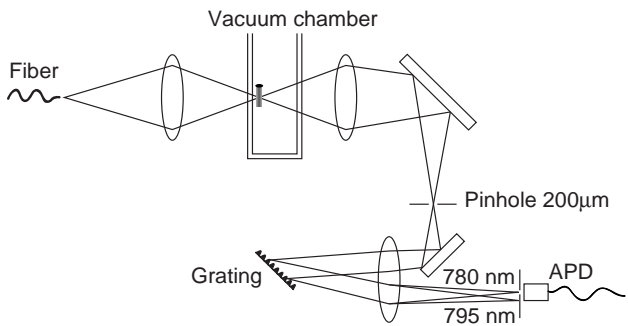,angle=0,width=\columnwidth}}
\caption{Schematic setup of the detection scheme. The Raman laser
light and the absorption laser light  are delivered by an optical
fiber to vacuum chamber and are focused  onto the atom laser beam
500\,$\mu$m below the Bose-Einstein condensate. The interaction
region between the atoms and the light fields is imaged onto a
pinhole, which spatially blocks light that has not illuminated the
atom laser beam. Subsequently, the absorption signal is separated
from the Raman laser light field by a grating spectrometer and
detected by an avalanche photo diode (APD).} \label{fig2}
\end{figure}

The laser light for the Raman transition and for the absorption
detection is generated by grating stabilized diode lasers and is
guided by a single mode optical fiber to the vacuum chamber (see
Fig. \ref{fig2}). The interaction region between the atoms and
the laser beams is imaged by a lens onto a pinhole of 200\,$\mu$m
diameter. The magnification of the imaging system is 4 so that
the transverse size of the atom laser beam matches that of the
pinhole and laser light which has not illuminated the atom laser
beam is blocked. After the pinhole the laser beams are collimated
and the different wavelengths are separated by a grating
spectrometer in Littrow configuration. The difference in
wavelength between the Raman laser light on one side and the
absorption laser light on the other side is 15\,nm which can
easily be resolved using a grating with 1800 lines/mm. The
diffraction efficiency into the 1st order is better than 80\%.
The absorption laser light (780\,nm) is detected by an avalanche
photo diode (APD), which is cooled to $-20^\circ$C and the signal
is recorded by a digitizing oscilloscope.

The intensity of the absorption light is chosen to be 1/5  of the
saturation intensity of 1.6\,mW/cm$^2$ to prevent saturation of
the optical transition. The laser power incident onto the atom
laser beam is about 5\,nW. This low light level requires very high
gain in the detection electronics. The avalanche photo diode has
an internal gain of more than 100\,A/W and the photo current is
subsequently amplified by low-noise amplifiers. The measured RMS
noise corresponds to 5\,nA photo current, which is a factor five
larger than the specified dark current of the avalanche photo
diode. With a reduction of the bandwidth of the electronics
(currently about 10\,kHz) one might approach the dark current
noise limit. The signal-to-noise ratio may also be improved by
intensity stabilizing the absorption laser or by making a
differential measurement by comparing the absorption signal with
the time dependent intensity of the absorption laser beam using
two identical photo diodes.

\begin{figure}
\centerline{\psfig{file=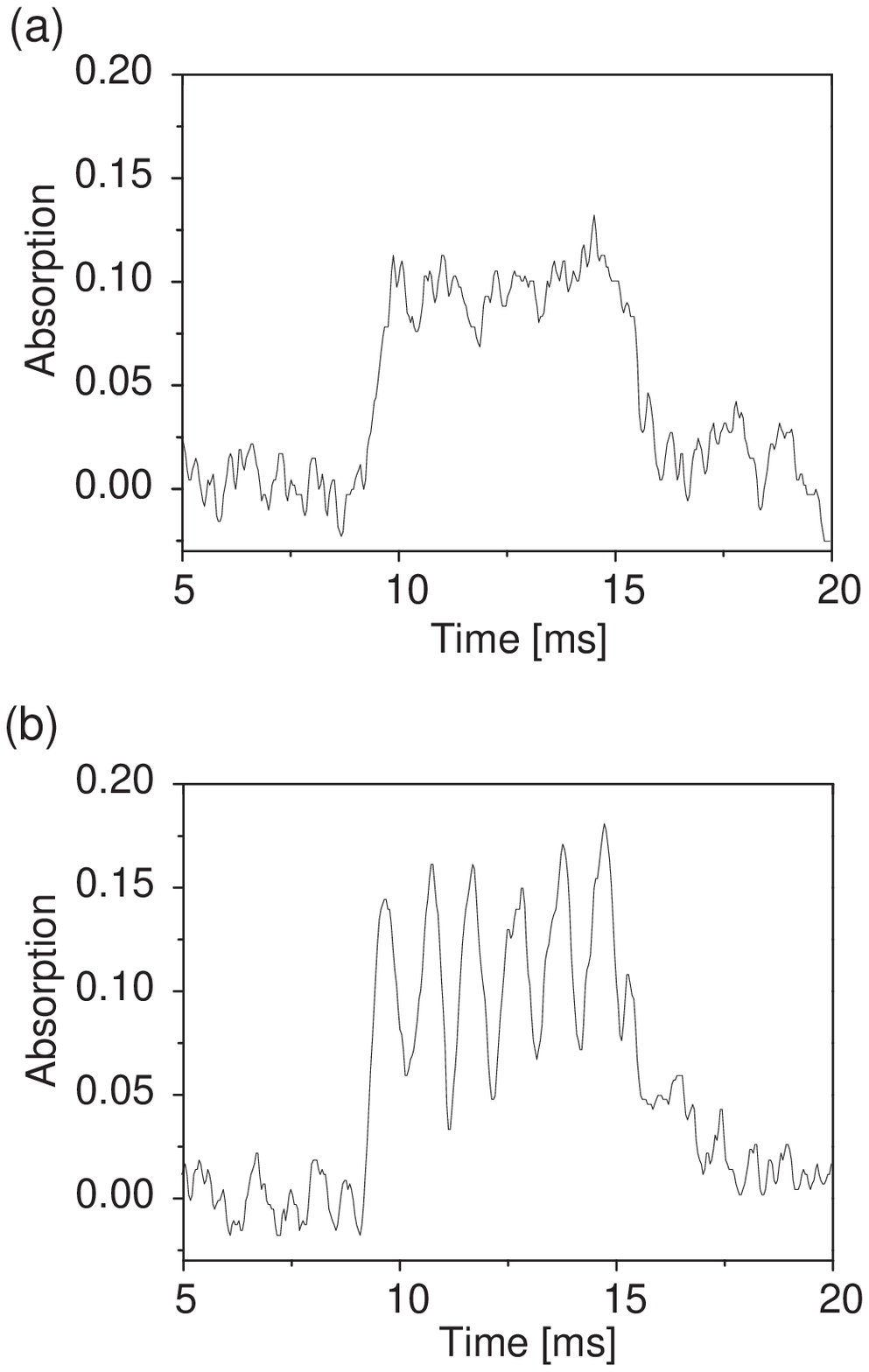,angle=0,width=0.7 \columnwidth}}
\caption{Absorption signals of atom laser beams. {\bf (a)} Signal
of an atom laser extracted for a duration of 6\,ms from the
condensate. The origin of time axis is at the beginning of the
output coupling. {\bf (b)} Absorption trace of two interfering
atom laser beams. The energy difference of 1\,kHz results in an
intensity modulation with a periodicity of 1\,ms. The RMS noise
of $1.1\times 10^{-2}$ is a factor of five larger than the
specified dark current noise of the photo diode and can be
attributed to electronic noise of the amplifier and intensity
noise of the detection laser.} \label{fig3}
\end{figure}

Fig. \ref{fig3} shows the data of absorption traces of the atom
laser beam.  The signal of an atom laser beam which is extracted
for a duration 6\,ms from a Bose-Einstein condensate is shown in
Fig. \ref{fig3}a. The time axis is counted from the beginning of
the output coupling.

A modulated  density of the atom laser beam is obtained, when the
output coupling is performed with two different radio
frequencies. The two extracted atom laser beams interfere and the
fringe pattern depends on the difference between the radio
frequencies.  Such interference pattern have been observed in the
spatial domain \cite{Bloch2000a} where the the fringe spacing is
position dependent due to the acceleration of the atoms in the
gravitational field. However, in the time domain the fringe
spacing is given by the frequency difference of the RF fields.
Figure \ref{fig3}b shows the absorption signal of two interfering
atom laser beams with an energy difference of 1\,kHz. This
results in the temporal modulation of the absorption signal with
1\,ms periodicity.

The presented technique is non-destructive for the Bose-Einstein
condensate and those parts of the atom laser beam that have not
yet reached the detection region. Both, the condensate and the
atom laser beam are formed in the $F=1$ hyperfine ground state.
With respect to this state the Raman lasers are detuned 70\,GHz
from the 5S$_{1/2}$\,$\rightarrow $\,5P$_{1/2}$ single photon
resonance and the very weak absorption light is detuned 6.8\,GHz
from the 5S$_{1/2}$\,$\rightarrow$\,5P$_{3/2}$ single photon
resonance, i.e.~all light fields are far detuned and spontaneous
scattering is suppressed. In contrast, optical pumping always
requires scattering of resonant photons which imposes a great
risk of an unwanted heating the sample by stray light.

The temporal resolution of the detection scheme is determined by
the transit time of the atoms through the interaction region. For
a dropping distance of 500\,$\mu$m and a waist of the absorption
laser beam of 26\,$\mu$m in the vertical direction, the
resolution is about 260\,$\mu$s. The temporal resolution may be
increased by either employing faster atoms (e.g.~after a larger
dropping distance or after acceleration by magnetic forces) or by
focusing the absorption laser to a smaller waist. Obtaining a
temporal resolution of a few ten microseconds appears feasible
with this scheme.

The continuous detection allows for the investigation of
processes with long time constants. For example, we could for the
first time monitor atom laser output coupling for up to 300\,ms.
Since the energy width of an atom laser is dependent on the
output coupling duration \cite{Koehl2001a}, with long extraction
times one might approach energy widths of a few Hz, where phase
diffusion processes in the condensate become observable
\cite{Graham1998a}. Another application could be the online
monitoring of the growth of a Bose-Einstein condensate. The
formation takes place on the time scale of a few hundred
milliseconds \cite{Miesner1998a,Robert2001a,Koehl2001b} and the
increase of atomic density will clearly show up in the atom laser
signal. The presented method may also prove to be a crucial tool
for investigations of continuous atom laser schemes.

In conclusion, we have presented a scheme to detect atom laser
beams continuously. The scheme is non-destructive to the
Bose-Einstein condensate in the trap and to those parts of the
atom laser beam, which have not yet entered the detection region.
Continuous atom laser operation for 300\,ms has been observed and
the data are obtained in real-time, which might enable experiments
with feedback.

We would like to thank A. Scheich for assistance with the photo
diode, and DFG for financial support.

\begin{references}
\bibitem{atomlasers} M. -O. Mewes {\it et al.}, Phys. Rev. Lett. {\bf 78}, 582 (1997);
B. P. Anderson and M. A. Kasevich, Science {\bf 282}, 1686
(1998); E. W. Hagley  {\it et. al.}, Science {\bf283}, 1706
(1999); I. Bloch, T. W. H\"ansch, and T. Esslinger, Phys. Rev.
Lett. {\bf82}, 3008 (1999).
\bibitem{Koehl2001a} M. K\"ohl, T. W. H\"ansch, and T.
Esslinger, Phys. Rev. Lett., in press (2001).
\bibitem{Lecoq2001a} Y.~ Le Coq {\it et al.}, cond-mat/0107032.
\bibitem{Andrews1996a} M. R. Andrews {\it et al.}, Science {\bf 273}, 84
(1996).
\bibitem{Bloch1999a} I. Bloch, T. W. H\"ansch, and T.
Esslinger, Phys. Rev. Lett. {\bf82}, 3008 (1999); T. Esslinger,
I. Bloch, and T. W. H\"ansch, Laser Spectroscopy XIV, ed. R.
Blatt, J. Eschner, D. Leibfried, F. Schmidt-Kaler, Singapore
(1999).
\bibitem{Esslinger1998a} T. Esslinger, I. Bloch, and T. W.
H\"ansch, Phys. Rev. A {\bf58}, R2664 (1998).
\bibitem{Bloch2001a} I. Bloch, M. K\"ohl, M. Greiner, T. W.
H\"ansch, and T. Esslinger, Phys. Rev. Lett. {\bf 87}, 030401
(2001).
\bibitem{Bloch2000a} I. Bloch, T. W. H{\"a}nsch, and T.
Esslinger, Nature {\bf 403}, 166 (2000).
\bibitem{Graham1998a} R. Graham, Phys. Rev. Lett. {\bf 81}, 5262
(1998), and references therein.
\bibitem{Miesner1998a}  H. J. Miesner {\it et al.}, Science {\bf 279},
1005 (1998).
\bibitem{Robert2001a} A. Robert {\it et al.}, Science {\bf 292},
461 (2001).
\bibitem{Koehl2001b} M. K\"ohl, T. W. H\"ansch, and T.
Esslinger, cond-mat/0106642.



\end{references}
\end{document}